\documentclass[prl,twocolumn,aps,english,showpacs,longbibliography,superscriptaddress]{revtex4-2}
\usepackage{natbib}

\usepackage{color}
\def\mathcolor#1#{\@mathcolor{#1}}
\def\@mathcolor#1#2#3{%
  \protect\leavevmode
  \begingroup\color#1{#2}#3\endgroup
}

\usepackage{graphicx}
\usepackage{dcolumn}
\usepackage{bm}
\usepackage{color}
\usepackage{tabularx}
\usepackage{amsmath,amsfonts,amsthm,amssymb}
\usepackage{appendix}
\usepackage[english]{babel}
\usepackage{mathptmx}
\usepackage{graphicx}
\usepackage{xcolor}

\begin{document}

\title{A route to minimally dissipative switching in magnets via THz phonon pumping}
\author{Mara Strungaru}
\email{mara.strungaru@york.ac.uk}
\affiliation{School of Physics, Engineering and Technology, University of York, York, YO10 5DD United Kingdom}

\author{Matthew~O~A~Ellis}
\affiliation{Department of Computer Science, University of Sheffield, Sheffield, S1 4DP, United Kingdom} 

\author{Sergiu~Ruta}
\affiliation{College of Business, Technology and Engineering, Sheffield Hallam University, Sheffield, S1 1WB, United Kingdom}

\author{Richard~F~L~Evans}
\affiliation{School of Physics, Engineering and Technology, University of York, York, YO10 5DD United Kingdom}

\author{Roy~W~Chantrell}
\affiliation{School of Physics, Engineering and Technology, University of York, York, YO10 5DD United Kingdom}

\author{Oksana Chubykalo-Fesenko}
\email{oksana@icmm.csic.es}
\affiliation{Instituto de Ciencia de Materiales de Madrid, CSIC, Cantoblanco, 28049 Madrid, Spain}

\begin{abstract} 
%Advanced magnetic recording paradigms typically use large temperature changes to drive switching which is detrimental to device longevity, hence finding non-thermal routes is crucial for future applications. By employing atomistic spin-lattice dynamics simulations, we show efficient coherent magnetisation switching triggered by THz phonon excitation in insulating single species materials. The key ingredient is excitation near the $P$-point of the spectrum in conditions where spins typically cannot be excited and when manifold $k$ phonon modes are accessible at the same frequency. Our model predicts the necessary ingredients for low-dissipative switching  and provides new insight into THz-excited spin dynamics. 
Ultrafast switching of magnetic materials has been shown to be predominantly thermally driven, but excess heating limits the energy efficiency of this process. By employing atomistic spin-lattice dynamics simulations, we show that efficient coherent magnetisation switching of an insulating magnet can be triggered by a THz pulse. We find that switching is driven by excitation near the $P$-point of the phonon spectrum in conditions where spins typically cannot be excited and when manifold $k$ phonon modes are accessible at the same frequency. Our model determines the necessary ingredients for low-dissipative switching  and provides new insight into THz-excited spin dynamics with a route to energy efficient ultrafast devices.
\end{abstract}

\maketitle

\date{March 2020}

\maketitle

The control of magnetic order on an ultrafast timescale in an efficient and robust manner is crucial for the development of the next-generation of magnetic devices\cite{vedmedenko2020}. The interaction of magnetic materials with femtosecond laser pulses has shown multiple fundamental effects that culminate in the ability to switch the magnetisation by means of purely optical excitation~\cite{StanciuPRL2007,Radu2015}. However, in metallic systems this is accompanied by a large rapid temperature increase, which is detrimental to the long-term usage of device. The use of insulators can be beneficial in this respect, however, the routes for energy efficient switching, involving minimal energy losses, still needs to be found.

A number of new possibilities have  been presented recently using  ultrafast excitations in the terahertz (THz) regime by means of femtosecond lasers or THz sources \cite{kampfrath2011coherent}. One of the fundamental questions in this research is understanding the  angular momentum transfer between spins and lattice, with recent results demonstrating that angular momentum transfer between both systems primarily takes place on an ultrashort timescale (around 200fs) ~\cite{Dornes2019}, in contrast with previously considered timescales in the range of 100 ps~\cite{Beaujouan2012}. Therefore, at the picosecond timescale and below, the dynamics of spin and lattice occurs simultaneously and one system can excite the other. Firstly, excitation of the lattice can change magnetic parameters such as exchange~\cite{maehrlein2018dissecting, Melnikov2003,afanasiev2021ultrafast} or anisotropy~\cite{afanasiev2014} and therefore can excite magnetisation dynamics. Reciprocally, ultrafast excitation of the spin system can produce excitation of the lattice at the same timescale. An example of that is the ultrafast Einstein-de Haas effect \cite{Dornes2019} or localised spin-Peltier effect in antiferromagnets \cite{otxoa2020}.

The most exciting results, however, are related to the possibility of magnetisation switching via THz phonon pumping. Such a possibility was anticipated by theoretical investigations ~\cite{vlasov_2020,Kovalenko2013} based on a phenomenological model which includes magneto-elastic anisotropy.
Recent experiments by Stupakiewicz \textit{et al}~\cite{stupakiewicz2020ultrafast_nat} demonstrated ultrafast magnetisation  switching in the magnetic insulator YIG by means of resonant pumping of specific longitudinal optical phonon modes. The switching was explained by excitations of local stresses that induce magneto-elastic anisotropy. The results also suggest a new universal ultrafast switching mechanism which may be applied to a wide range of materials. 
 
Thus, energy-efficient magnetisation switching is particularly interesting to explore in insulators. This is due to the fact that, firstly, the electronic system with its low specific heat can have little participation in the energy uptake and, secondly, that excitation of the spin-phonon system on the sub-picosecond timescale will have minimal interaction with the outside world in terms of energy diffusion. Therefore, one can try to find conditions for almost dissipationless (``cold'') switching when the angular momentum is efficiently transferred from phonons to spin and the energy is used to switch magnetisation without losses.

\begin{figure}[htb!]      
  \includegraphics[trim={0cm, 0cm, 0cm, 0cm},width=0.75\linewidth]{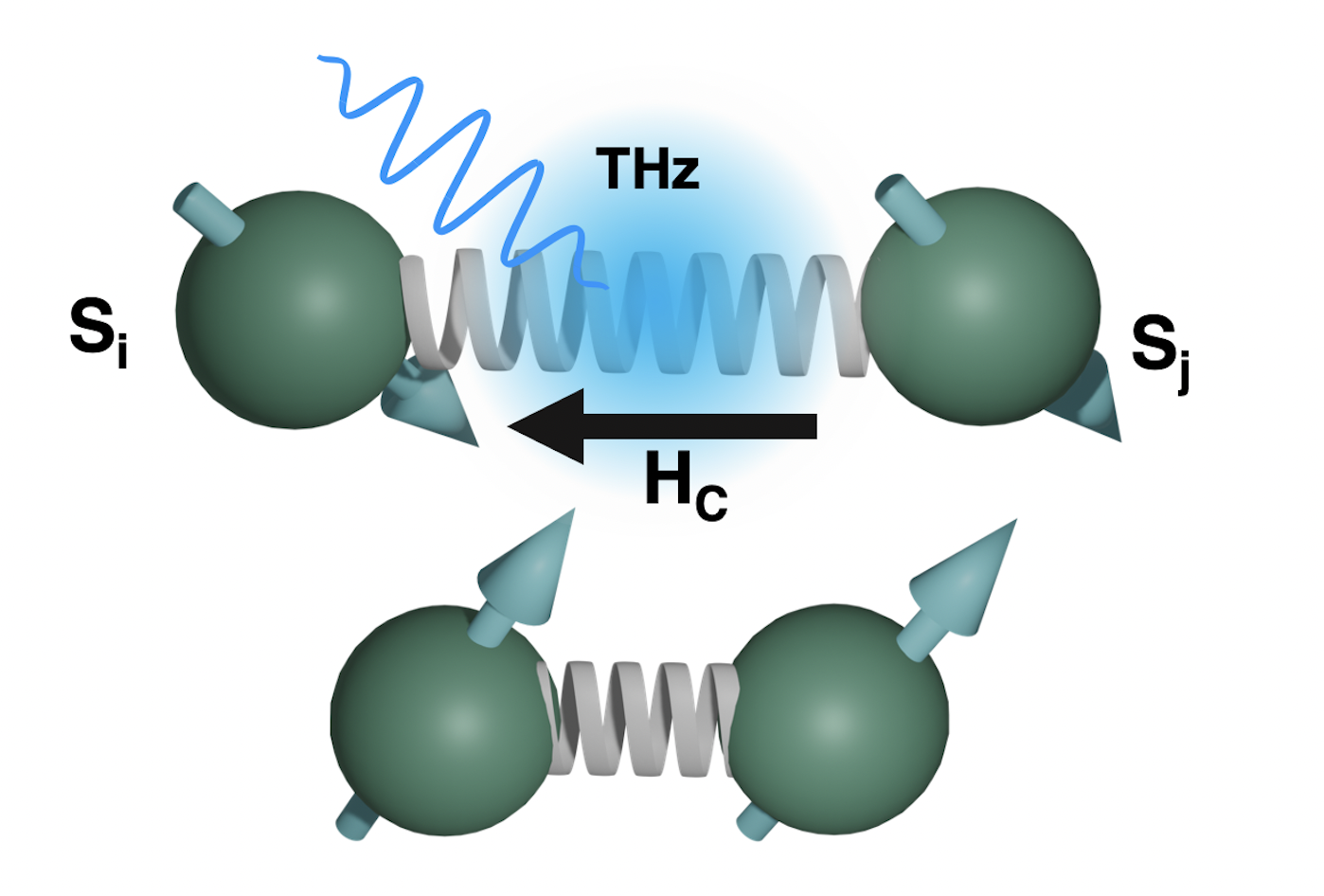}
\caption{Illustration of magnetisation switching by THz pulses. The THz pulse couples directly to the lattice creating a magneto-elastic field that leads to low dissipation switching. }
\label{illustration}
\end{figure}

In this Letter, we demonstrate magnetisation switching under the application of a THz phonon pulse, as shown schematically in Fig.~\ref{illustration}. The modeling is done within the molecular dynamics approach in a self-consistent spin-lattice framework~\cite{strungaru_2021_sld}. We demonstrate the possibility of an energy efficient switching in the conditions when phonons are excited with high $k$-values and THz frequencies, corresponding to a maximum in the density of states with no available spin-wave modes. The mechanism of switching is via local magneto-elastic fields created by atomic displacements due to the phonon pulse. In this condition, the absence of magnon excitations means that practically all phonon angular momentum is transferred to precessional magnetisation switching. The mK spin temperature developed during the switching process shows that the switching process can be considered non-dissipative.

The computational model for this work has been previously employed in the systematic investigation of equilibrium and dynamic properties of \textsc{BCC} Fe \cite{strungaru_2021_sld}. The Fe parameters were chosen as they are well studied in the literature and good parameterisation exists from theory and experiments, providing a realistic spin-phonon spectrum. However, for the sake of proof of concept, the system is effectively treated as a insulator - no electronic damping is considered on the spin system and the only thermostat is applied to the lattice. The system size used in the simulations is $10 \times 10 \times 10$ \textsc{BCC} unit cells. For phonon interactions, we consider a Harmonic Potential (HP) and  Morse Potential (MP). The HP is defined as $U(r_{ij})={V_0} (r_{ij}-r_{ij}^0)^2$, where $V_0$ has been parametrised for \textsc{bcc} Fe in \cite{Assmann2019} ($V_0=0.15eV/\text{\r{A}}^2$).  
The MP is defined as $U(r_{ij})=D [e^{ -2 a (r_{ij}-r_0)}- 2 e^{-a (r_{ij}-r_0)}]$ and is parameterized in \cite{girifalco1959application} for \textsc{bcc} Fe ($D=0.4174 eV, a=1.3885 \text{\r{A}}, r_0=2.845 \text{\r{A}}$). For both potentials the interaction range was restricted to $r_c=7.8\text{\r{A}}$. Magnon properties, such as damping and equilibrium magnetisation have been shown not to be influenced by the choice of potential~\cite{strungaru_2021_sld}. The spin Hamiltonian used in our simulations consists of contributions from the exchange interaction, Zeeman and anisotropy energies as described in~\cite{strungaru_2021_sld}. In our model system, we used an uniaxial anisotropy with the easy axis $\mathbf{e}$ parallel to z-direction to mimic the switching experiments. Finally, the spin-lattice coupling term ($\mathcal{H}_{\mathrm{c}}$) was taken to have a pseudo-dipolar form $
\mathcal{H}_c= -\sum_{i,j} f({r}_{ij}) \left[(\textbf{S}_i \cdot \hat {\textbf{r}}_{ij}) (\textbf{S}_j \cdot \hat {\textbf{r}}_{ij}) - \frac {1}{3} \textbf{S}_i \cdot \textbf{S}_j \right]$.
 
 The magnitude of the interactions is assumed to decay as $f(\textbf{r}_{ij})=C J_0/r_{ij}^4$ as presented in~\cite{Assmann2019} with $C$ taken as a constant, for simplicity, measured relative to the exchange interactions $J_0$ ($CJ_0=0.452 eV \text{\r{A}}^4$). The coupling strength $C$ can be parameterised via the strain-dependent magneto-elastic anisotropy or via the value of the damping of magnon/phonon origin \cite{strungaru_2021_sld}. For the spin degrees of freedom we solve only the precessional Landau-Lifshitz equation: $\frac{\partial \mathbf{S}_i }{\partial t} = -{\gamma}\mathbf{S}_i \times \mathbf{H}_i $ without the Gilbert damping term, as this appears intrinsically in our model via the direct coupling to the lattice.

\begin{figure}[hbt!]
  \includegraphics[trim={0cm, 0.05cm, 0.05cm, 0cm},width=\linewidth]{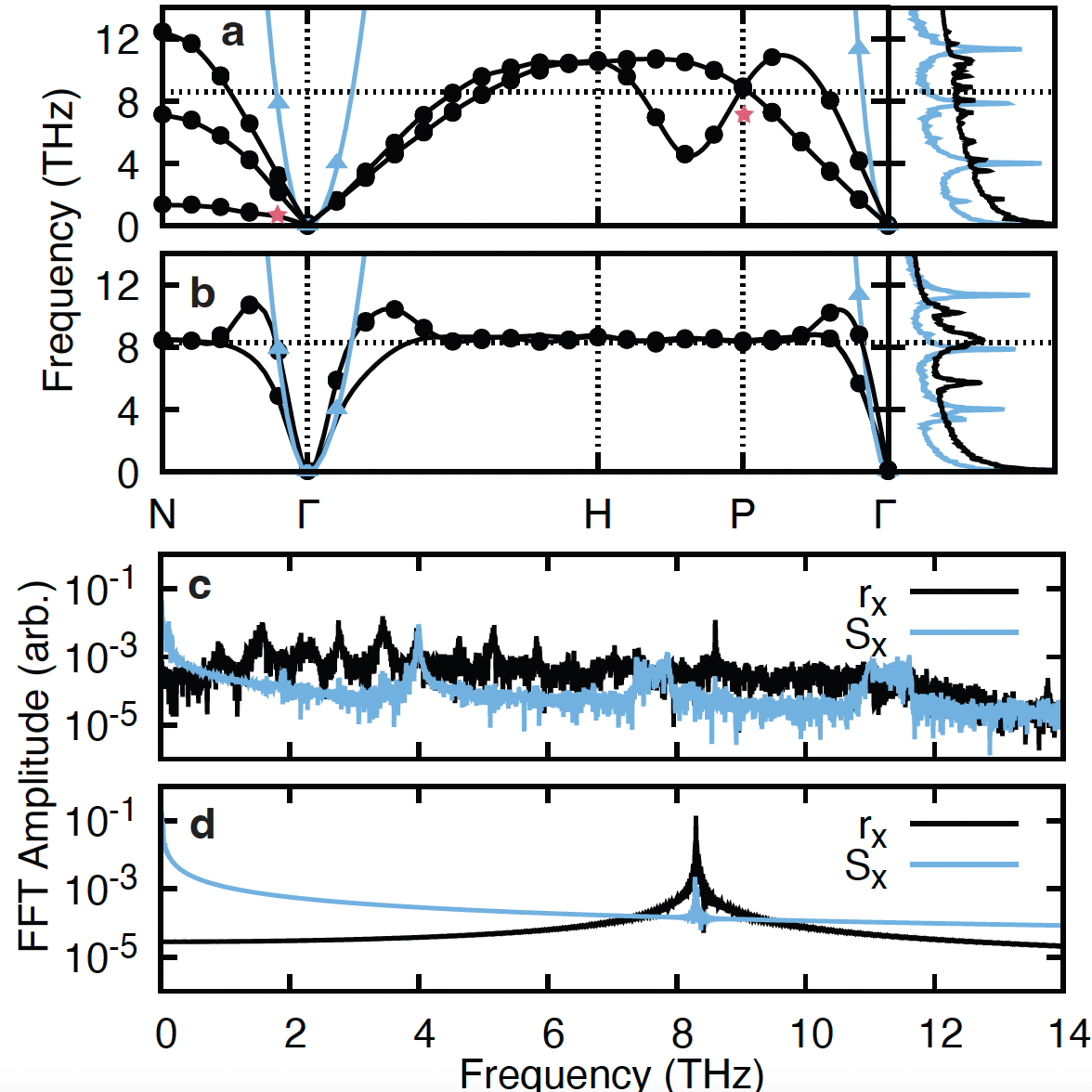}
\caption{Phonon (black points) and magnon (blue line points) spectra  and density of states (right panel, continuous lines) of the x component of velocities and spins, respectively  at T=10K for a Morse potential (panel a - MP)  and Harmonic potential (panel b - HP). The continuous black lines in the left figure in panels a, b represent the spectra calculated using all three components of velocities. The red symbols in panel a show where we excite the system (close to $\Gamma$ and in $P$ point).  The Fourier Transform of spin (magnons) and position (phonons) after excitation with  THz pulse for c) MP at $8.6$THz and d) HP at $8.3$THz.  }
\label{fig::spectrum}
\end{figure}

To model the effect of THz phonon excitation, we apply a periodic external force to each atom $ F^\alpha_{THz}(t, \mathbf{r})=f_0^\alpha \cos(2\pi \nu t + \mathbf{k} \cdot \mathbf{r}) \Theta(t_p-t)$, where $\Theta(t_p-t)$ is the Heaviside step function, $t_p$ is the rectangular pulse duration, $\alpha$ the $x, y, z$ coordinates of the forces. The excitation force $F^x_{THz}$ is non-homogeneous in space by choosing different $k$-vectors and for simplicity is applied along the $x$ direction. 

\begin{figure*}[htb!]     
  \includegraphics[trim={0cm, 0cm, 0cm, 0.1cm},width=\linewidth]{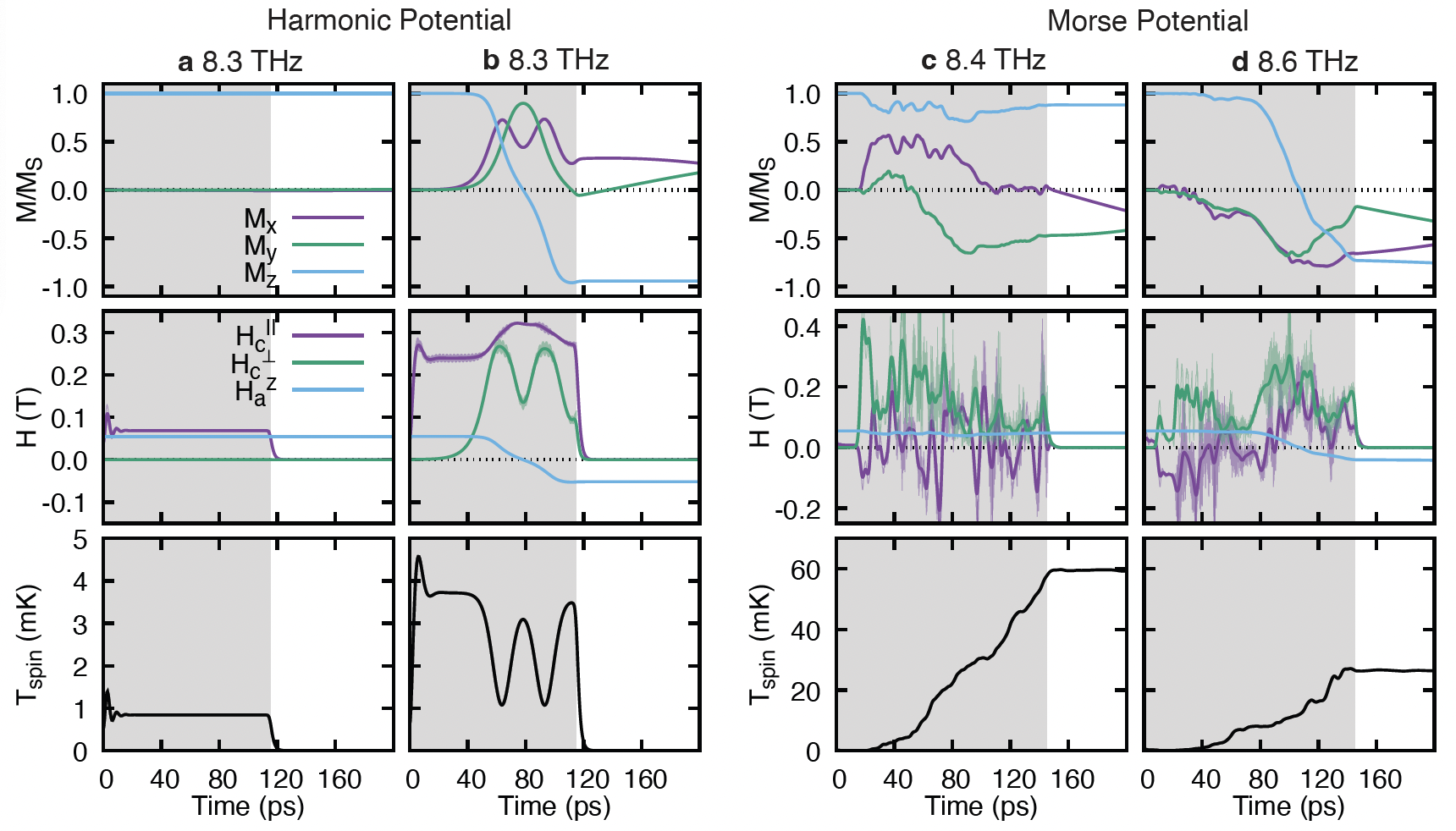}
  \caption{Evolution of magnetisation ($M/M_S$), coupling field parallel and perpendicular to the direction of magnetisation ($H_c^{\parallel}$ and  $H_c^{\bot}$), anisotropy field ($H_a^z$) and spin temperature ($T_{spin}$) under the application of a THz pulse for Harmonic (HP) and  Morse potential (MP). For the MP we also include the moving average for the perpendicular coupling field - orange curve.  The temporal region where the THz pulse is applied is emphasized by the gray area. The frequency and pulse width used in the calculations are: a) 8.3THz and 115ps, b)  8.2THz and 115ps, c) 8.6THz and 145ps,  d) 8.4THz and 145ps.}
{\label{fig::switching_magn_combined}}
\end{figure*}

The complete magnon-phonon spectrum is presented in Fig. \ref{fig::spectrum}. Note that for the HP, the phonon and magnon spectrum intersect, however the MP does not present this feature.
The power spectral density of the autocorrelation function in the frequency domain \cite{strungaru_2021_sld} (see also Fig. S1 in Supplementary Material) reveals a peak  for phonons at frequencies ca. 8.3 THz for the HP and a broad-band excitation for the MP around the similar frequency (with no available spin-wave modes), hence we first excite our systems at around 8THz. The application of the THz force drives the atom displacement to excite phonons within the system. Within the spin-lattice framework, these phonons break the local symmetry of the lattice which, through the pseudo-dipolar coupling term, generates an internal field capable of switching the magnetisation. The k-vector corresponding to the P-point (obelisk symbol in Fig. \ref{fig::spectrum}) and the application of the force on the x direction was selected as it gives rise to movement of the atoms entirely out of phase along the x-direction, with oscillations of the atoms around their equilibrium position up to $7\%$ of the lattice spacing (for $f_0^x=0.05$) - Supplementary Material, Fig. S2.
 In the case of the HP - Fig.\ref{fig::spectrum}, panel b), forced excitation with a THz pulse of 8.3~THz leads only to a response at the same frequency for both magnons (spin $S_x$) and phonons (position $r_x$) - Fig.\ref{fig::spectrum}, panel d). For the MP, although we excite the system only on the x direction and at the $P$ point in the Brillouin zone, we observe multiple spin and phonon modes ( Fig.\ref{fig::spectrum}, panel c) reflecting the decay of the forced excitation into other modes, along the  $P-\Gamma '$ path. This gives rise to displacements in all three directions, (Fig. S2, Supplementary Material) and a more complicated switching pattern and additional heating (as observed later in Fig. \ref{fig::switching_magn_combined}, \ref{fig::phase_diagram_T0K}).  Note that, the analysis of the coupling fields (see Supplementary Material, Fig. S1) shows a broadening of the spectra at the excitation frequency for the MP, in contrast with the peak observed for the HP, suggesting a  much weaker coupling to the THz force (hence a lower peak amplitude at the excitation frequency).

In spite of these differences in the excitation spectrum, we have observed magnetisation switching for both potentials. Fig. \ref{fig::switching_magn_combined} shows examples of switching cases for HP (panel a) and MP (panels c). We observe that the parallel to the magnetisation component of the coupling field $H_c^{\parallel}$ is zero, and the THz excitation leads to the apparition of a perpendicular component $H_c^{\bot}$, which will drive the switching. In the case of the HP - panel a, we observe that at around 40 ps there is an increase in the $H_c^{\bot}$, leading to the initiation of switching. For the case shown in panel b, the perpendicular field developed is smaller and cannot lead to switching. For MP, we observe that the perpendicular coupling field fluctuates due to the wider range of excited phonons. The switching is triggered when this field reaches a relatively constant, large value, which is a random event. For the case of the panel d, no switching is triggered since the in-plane component of the perpendicular field is rotating (see Fig. S5 in Supplemental Material). 
Importantly, after the THz excitation, the magneto-elastic fields are almost zero showing that all the energy was transferred and used during the switching. After this time, the only effective field that leads the magnetic system back to saturation is a small uniaxial anisotropy, $H_a^z=0.05$T, which defines a slow relaxation of the system at much longer timescales  ($>30$ns) than those presented here.

To characterize the spin system we calculate the spin temperature \cite{ma2010temperature}, see Fig.~\ref{fig::switching_magn_combined}, bottom panels. We observe that the increase of temperature is in the $mK$ range, proving the lack of heating during this process. Moreover, the spin temperature in the case of the HP is raised to several $mK$ while approximately 10 times larger temperatures (ca. 10 $mK$ and up to 28 $mK$) are created for the MP. In the case of the HP, the very small temperature increase is due to the fact that no magnons can be excited in the spin system and thus the energy goes efficiently into the magneto-elastic fields created by the spin-phonon coupling with the rest of the increase in temperature coming from switching only as shown in Fig.~S3, Supplementary Material. This underlines the main differences between the HP and an-harmonic MP. In the latter case, large non-linearities are present in the system.  Since our excitations are strong, these non-linearities act as an efficient scattering mechanism for phonons which finally add a temperature-like effect, with quite slow relaxation.

Fig. \ref{fig::phase_diagram_T0K} shows the corresponding phase diagram at $T=0$K for HP (a) and MP (b) in terms of the excitation frequency and the pulse duration. An important difference between the two potentials is a much larger switching region for the MP, however with scattered points due to scattered phonon modes and finite-size effects which we will discuss below. We note that the excitation at this frequency produces a large phonon response, up to 5-7\% of the interatomic distance (see Supplementary Material, Fig. S4) and many phonon modes are available at these frequencies. We underline that magnons do not have modes at this frequencies and corresponding k-point hence the switching is triggered by phonons. During the switching, the change in the magnetisation length is less than $0.04\%$, suggesting once more a non-thermal switching. 

One of the main characteristics of the switching is that a minimum duration of the THz pulse, ca. 50 ps is required, as shown by the phase diagram. A detailed examination shows that the switching is precessional due to the fact that the application of the force on the x direction for the k-vector corresponding to the $P$ point in the Brillouin zone was selected to generate a coupling field that acts in plane, on the x direction (see Fig. \ref{fig::switching_magn_combined}). Additional proof is shown in Fig. S6, Supplementary Material, where we impose an x displacement on atoms of the same magnitude, frequency and phase as in the case of the numerical simulations with a HP. This periodic motion creates a perpendicular coupling field to the magnetisation and leads to precessional switching.
Panels c) and d) in Fig.\ref{fig::phase_diagram_T0K} shows the final magnetisation state for different pulse widths (points) after the application of a THz laser pulse. For longer pulse widths, there is an oscillatory behavior of the magnetisation which corroborates the precessional and coherent nature of the switching.

\begin{figure}[htb!]    
  \includegraphics[trim={0cm, 2cm, 8cm, 0cm},width=1\linewidth]{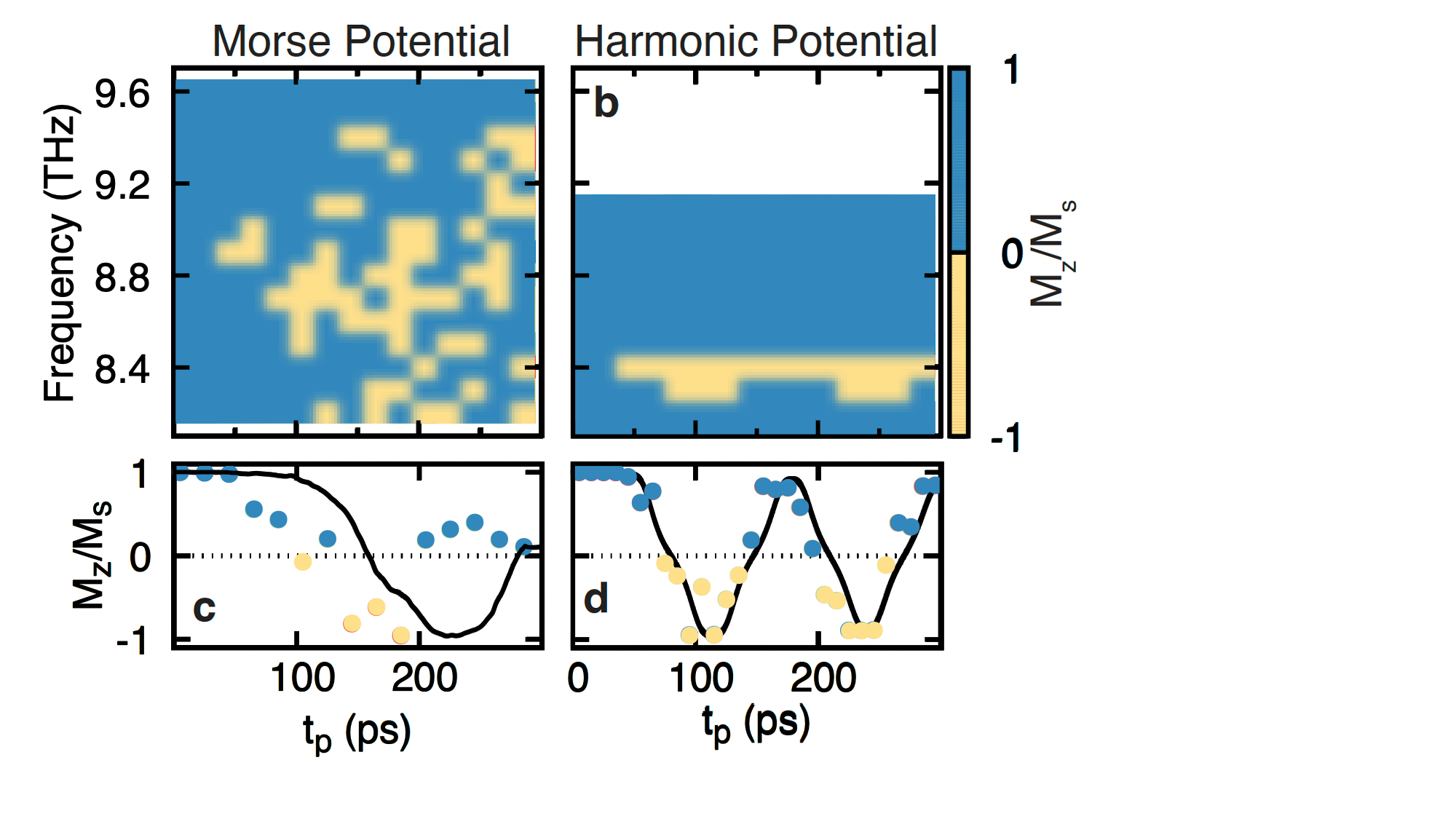}
\caption{Switching phase diagrams, $T=0$K as a function of THz pulse frequency and duration of the pulse, Morse (MP - panel a) and Harmonic (HP - panel b) potential. Panels c) and d) - final magnetisation after each THz pulse (points) and the magnetisation during 295 ps pulse (continous line).}
\label{fig::phase_diagram_T0K}
\end{figure}

The fluctuation of the in-plane coupling field for MP is responsible for random switching events visible in the phase diagram of Fig.\ref{fig::phase_diagram_T0K}, in contrast to the regular behavior of HP. The random switching effects are especially visible in small systems, as is the case of our simulations. Analysing panels c and d in Fig. \ref{fig::phase_diagram_T0K}, we confirm that in the case of the MP - panel c, the scattering of the phonon modes and subsequent heating leads to a scattered final magnetic state with each realisation, while for the HP, a similar magnetic state is obtained with each realisation. A similar ``randomization'' has been produced when we analysed the switching diagrams at non-zero temperatures for the HP, see Supplementary Material, Fig. S2. This ``random'' switching diagram has been also reported for a macrospin nanoparticle with magneto-elastic anisotropy term at non-zero temperature~\cite{vlasov_2020}. 

\begin{figure}[htb!] 
   \includegraphics[trim={0cm, 2.5cm, 17cm, 0cm},width=0.95\linewidth]{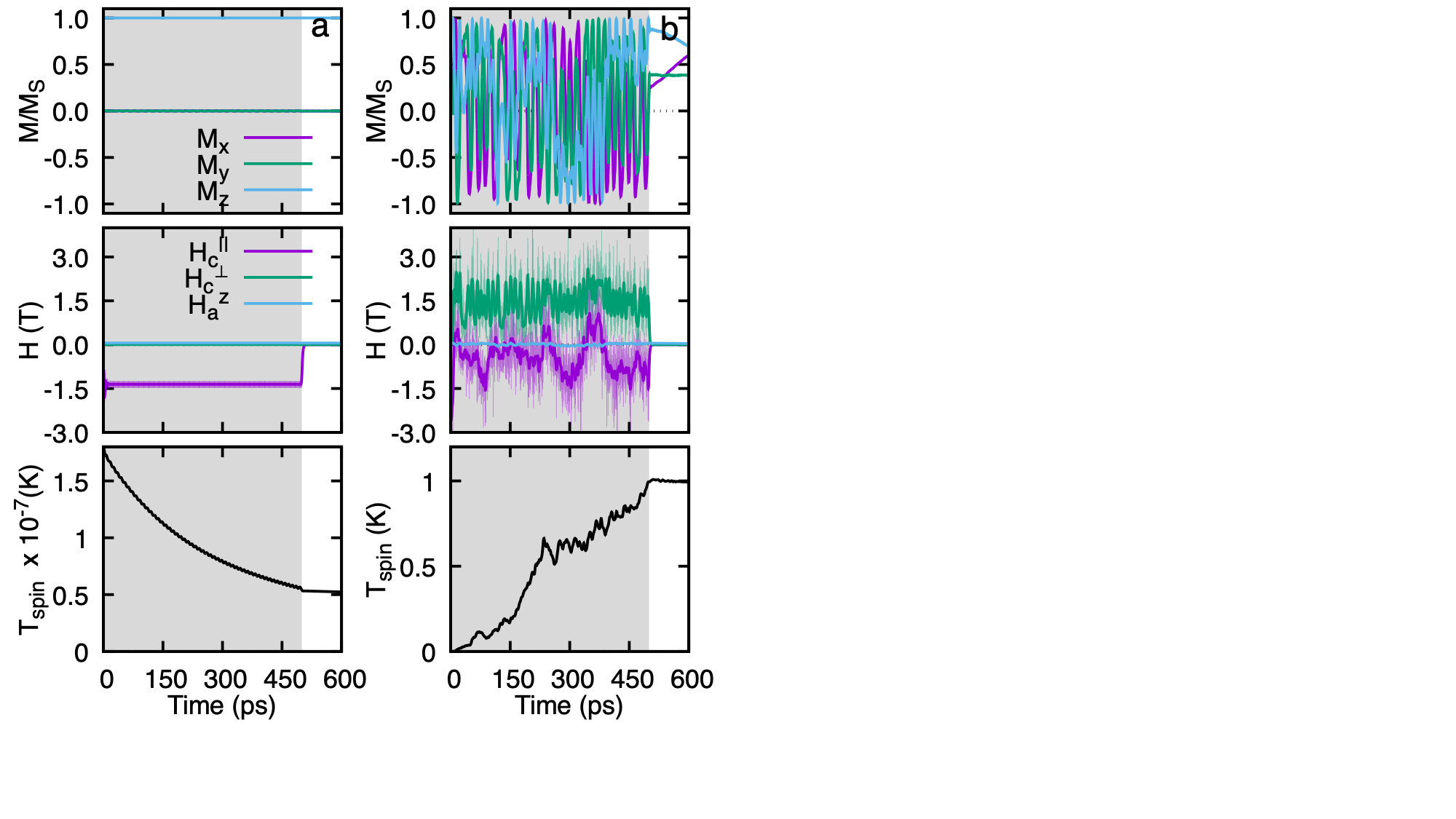}
  \caption{Example of dynamics excited near $\Gamma$-point at 0.4THz for Morse potential and two excitation strengths: $f_0^x=0.02$ (left), $f_0^x=0.03$(right).}
\label{fig::Gamma}
\end{figure}

Finally, Fig.\ref{fig::Gamma} presents the results of lower frequency THz-phonon excitation near the $\Gamma$-point (at the red asterix symbol in Fig. \ref{fig::spectrum}) for the MP and for two excitation strengths. This is a characteristic example of what typically occurs in both systems. A small excitation, $f_0^x=0.02$, does not develop any transverse coupling field and produces no effect except small-amplitude spinwaves. Larger excitations (for $f_0^x=0.03$, average $H_c^t=1.5$T)  randomises the spin system leading to a large increase of its temperature. Hence, although excitation close to the $\Gamma$-point produces random (non-coherent) switching events, for some particular excitation strengths, the process is associated with subsequent heating. The final spin temperature increases when the excitation strength increases.

In conclusion, using a spin-phonon model with angular momentum transfer, we investigated the effect of magnetisation switching driven by phonon excitations with minimal energy dissipation. SLD models are crucial for the investigation of magnetisation switching via THz phonons, as  we are able to access the magnetisation dynamics corresponding to the excitation of individual or collective phonon modes. Our results 
suggest that ferromagnetic materials that present a flat phonon spectral region where a large number of phonon modes can be efficiently excited, are good candidates for THz-assisted ``cold'' switching. The key factor is excitation with THz phonons with frequencies and $k$-points at a maximum in the phonon density of states and no spin excitations. The mechanism corresponds to the phonon-driven generation of magneto-elastic fields with components perpendicular to the  magnetisation producing precessional switching on the 100 ps timescale. 
Importantly, we predict this possibility for the case of single species materials unlike the first experimental observations~\cite{stupakiewicz2020ultrafast_nat} where the switching was due to optical phonon excitation. In this regard, the important factor is related to the possibility of experimental access to the excitation with $k$-vectors close to this ``flat'' phonon spectrum, in our case the $P$-point.
Our prediction may be of crucial importance for next-generations of eco-friendly storage devices since heat production is one of the major problems for large data storage centers.

\begin{acknowledgments}
 Financial support of the Advanced Storage Research Consortium and ARCHER2-eCSE06-6 is gratefully acknowledged. MOAE gratefully acknowledges support in part from EPSRC through grant EP/S009647/1. The simulations were undertaken on the VIKING cluster, at the University of York. SR, RWC, RFLE acknowledges funding from European Union's Horizon 2020 Research and Innovation Programme under grant agreement No. 737709.  The authors acknowledge the networking opportunities provided by the European COST Action CA17123 "Magnetofon" and  the short-time scientific mission awarded to  MS.
 \end{acknowledgments}

\bibliography{references}

\vfill
\clearpage
\newpage
\section{Supplementary information}
\setcounter{figure}{0}
\setcounter{section}{0}
\setcounter{equation}{0}
\renewcommand\thefigure{S\arabic{figure}}
 Fig. \ref{fig::power_spectra_coupling_morse_harmonic} shows the power spectrum of magnons ($S_x$), phonons ($v_x$) and coupling field ($H^c_x$) in the case of Morse Potential (MP) and Harmonic Potential (HP). In the case of HP, since the phonons present a large peak due to the flat phonon spectra, the coupling field will inherit  this peak and will lead to a strong magnon-phonon coupling which translates to a strong spin response when applying the THz pulse. This is not the case with the Morse potential, where the coupling has lower intensity peaks, and hence, the magnons are not so efficiently coupled to the THz pulse.

\begin{figure}[htb!]
  \includegraphics[trim={3cm, 7cm, 9cm, 0cm},width=0.85\linewidth]{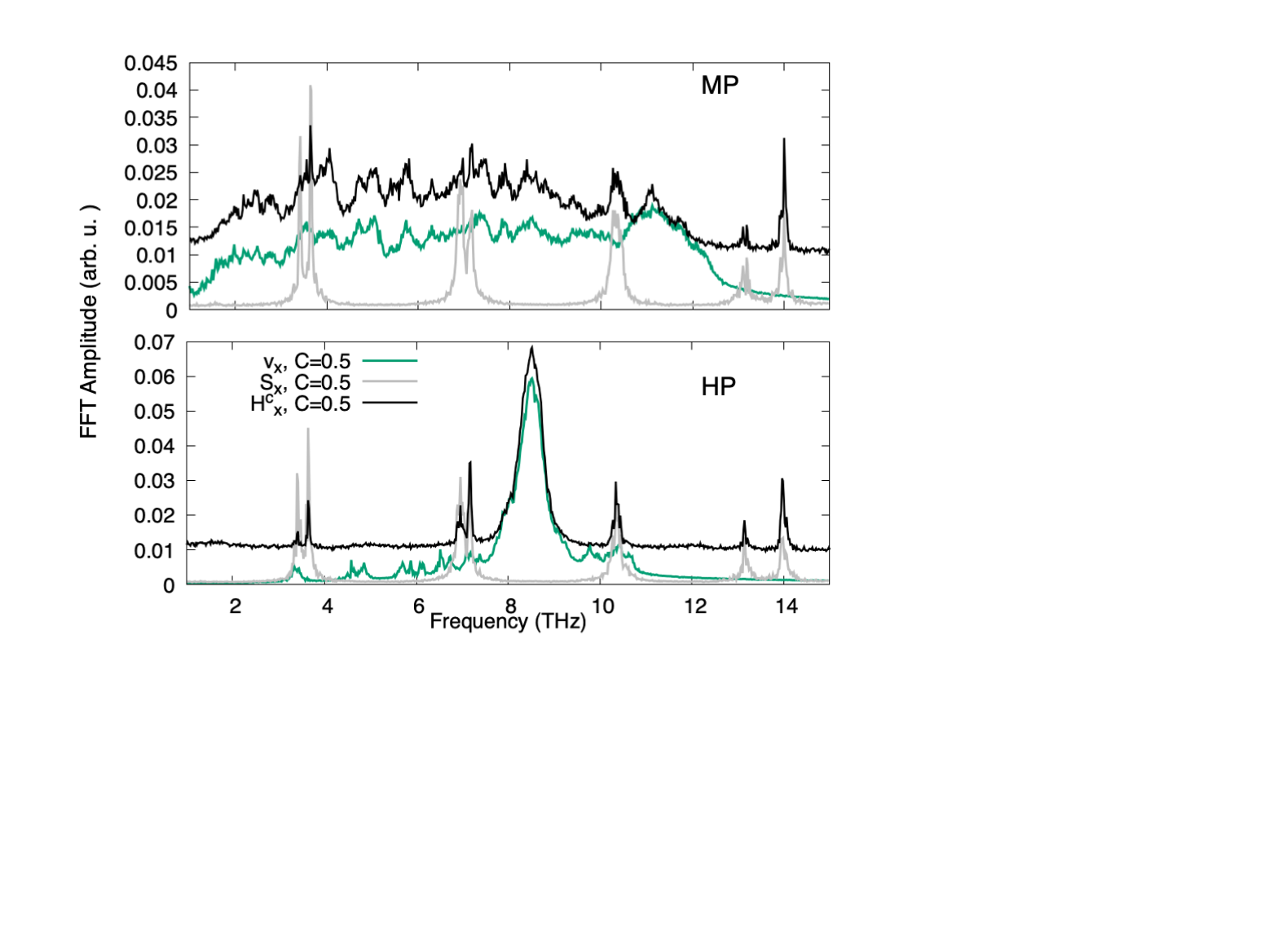}
  \caption{The power spectral density of the autocorrelation function in the frequency domain for magnons, phonons, and coupling field for a SLD simulations with a Morse (top panel) and Harmonic potential (bottom panel).}
\label{fig::power_spectra_coupling_morse_harmonic}
\end{figure}

\begin{figure}[htb!]
  \includegraphics[trim={0cm, 11cm, 17cm, 0cm},width=0.85\linewidth]{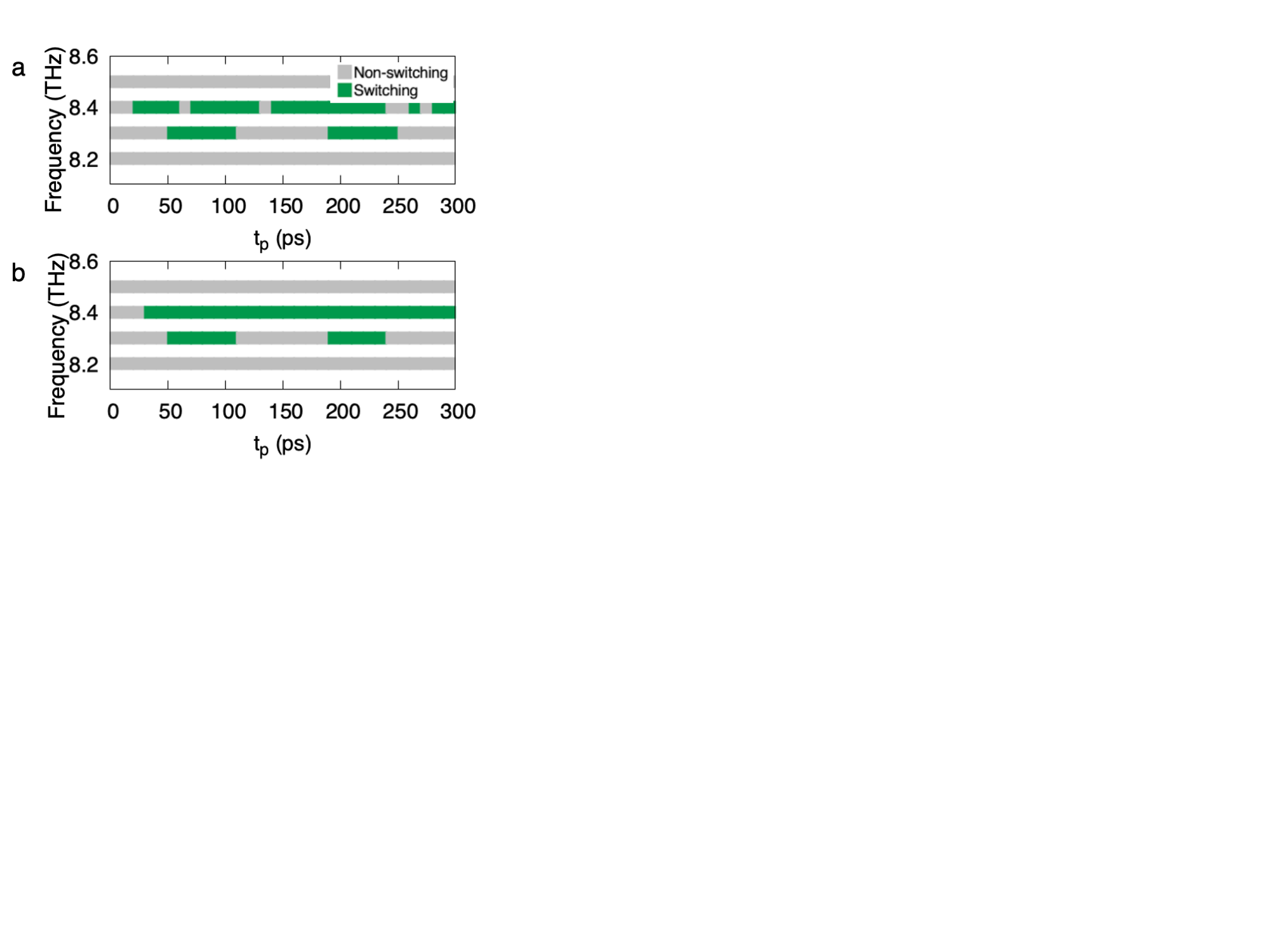}
  \caption{Phase diagram, harmonic potential, T=10K. Top panel - one random realisation; bottom panel - 4 random realisations;}
  \label{phase_diagram_t_10}
\end{figure}
 Fig. \ref{phase_diagram_t_10} shows the phase diagram of THz switching at finite temperature, $T=10K$ in the case of Harmonic potential. The temperature effect leads to a randomisation of the phase diagram (panel a) in comparison with 0K results.To reduce the thermal noise, the switching phase diagram is averaged over multiple
realisations - panel b). We observe that switching occurs in the same frequency range as for the $T=0K$ ( Fig. \ref{fig::phase_diagram_T0K}), however the initial pulse width necessary for switching decreases with approximately 10ps. The decrease
can be due to the fact that phase diagram is averaged only for 4 realisation, hence a small
change like this of the pulse width can still exhibit an effect of the thermal fluctuations.

\begin{figure}[htb!]
  \includegraphics[width=0.85\linewidth]{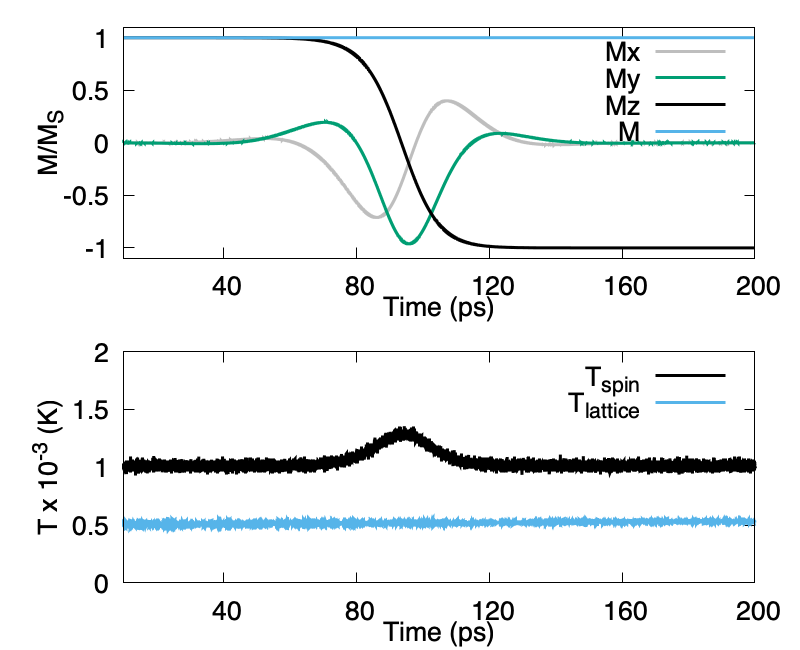}
  \caption{Magnetisation switching in an applied field of -1T, spin damping 1.0 and no lattice damping. Top panel - evolution of magnetisation and bottom panel - evolution of spin and lattice temperature; }
\label{fig::switching_constant_field}
\end{figure}

To investigate how much heating occurs during magnetisation switching, in Fig. \ref{fig::switching_constant_field} we show the increase in the spin temperature when applying a magnetic fields of $-1T$. We observe that the increase in the spin temperature is of about 0.5mK, the same order of magnitude as the increase in temperature when switching occurs via THz excitation, in the case of Harmonic Potential. Since the spin system is coupled to a thermostat in this case (so it will allow for fast switching dynamics) after the switching, the system goes to the temperature of the thermostat of 1K.

\begin{figure}[!htb]\centering
  \includegraphics[trim={0cm, 10cm, 16cm, 0cm},width=0.95\linewidth]{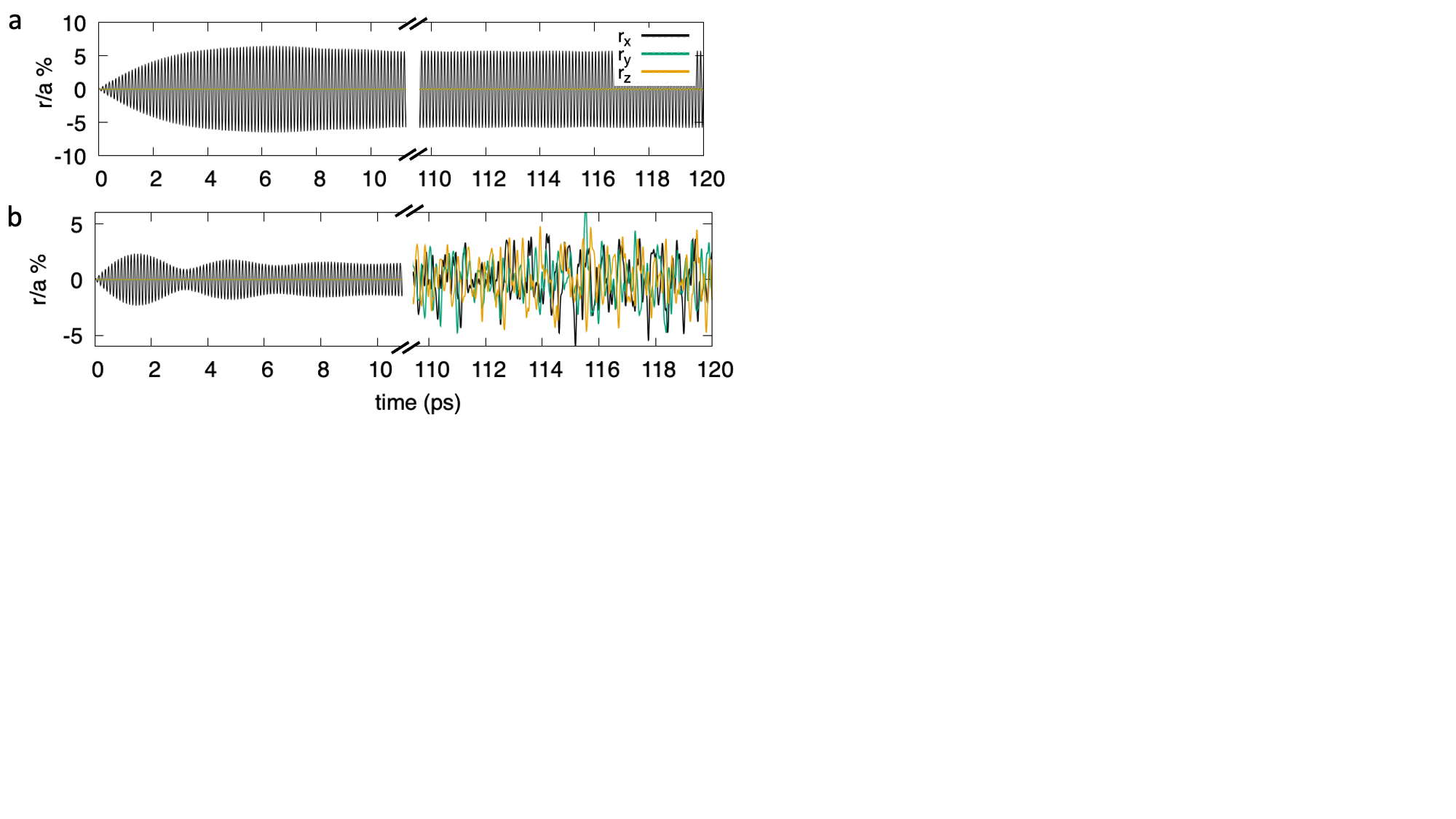}
  \caption{Temporal evolution of the displacements of the atom at position (0,0,0) normalised to the lattice constant for a Harmonic (panel a) and Morse potential (panel b). The THz frequencies are 8.3THz for the Harmonic potential and 8.6THz for the Morse potential. The THz pulse is applied continuously during the simulation.}{\label{fig::displacements_and_fft}}
\end{figure}

During the application of the THz pulse, the lattice displacements are relatively low (less than $5-7\%$ depending on frequency of the THz pulse). Fig. \ref{fig::displacements_and_fft} shows the evolution of the displacements. For the Harmonic potential, excitation around the P point in the Brillouin zone (obelisk symbol in Fig. \ref{fig::spectrum}
leads to a periodic motion of the atoms on x direction (with no displacements on y and z coordinates). This gives rise to the FFT peak observed in Fig.\ref{fig::spectrum}, panel d). For the Morse potential (Fig. \ref{fig::displacements_and_fft}, panel b), although initially the displacements are on x direction, with the frequency corresponding to the one given by the THz excitation, this mode rapidly decays in phonon modes along the $P-\Gamma '$ path, as shown by the FFT of the x displacement in Fig.\ref{fig::spectrum}, panel c. The decay of the phonon modes will lead to a coupling field that strongly fluctuates in time, leading to scattered switching phase diagram.

The  perpendicular xyz components of the coupling field in the case of Morse potential (switching and non-switching events presented in panels c, d in Fig. \ref{fig::switching_magn_combined}) are shown below. For the non-switching case - panel b in Fig.\ref{fig::perp_comp_field} we observe that the coupling field fluctuates around zero. This suggests that the coupling field is rotating, and hence is not sufficient to trigger switching. Panel a, that corresponds to the switching case, still presents fluctuations, but after 50ps, when switching is triggered, the in-plane fluctuations are around a finite field value.

\begin{figure}[!htb]\centering
  \includegraphics[trim={0cm, 0cm, 0cm, 0cm},width=0.75\linewidth]{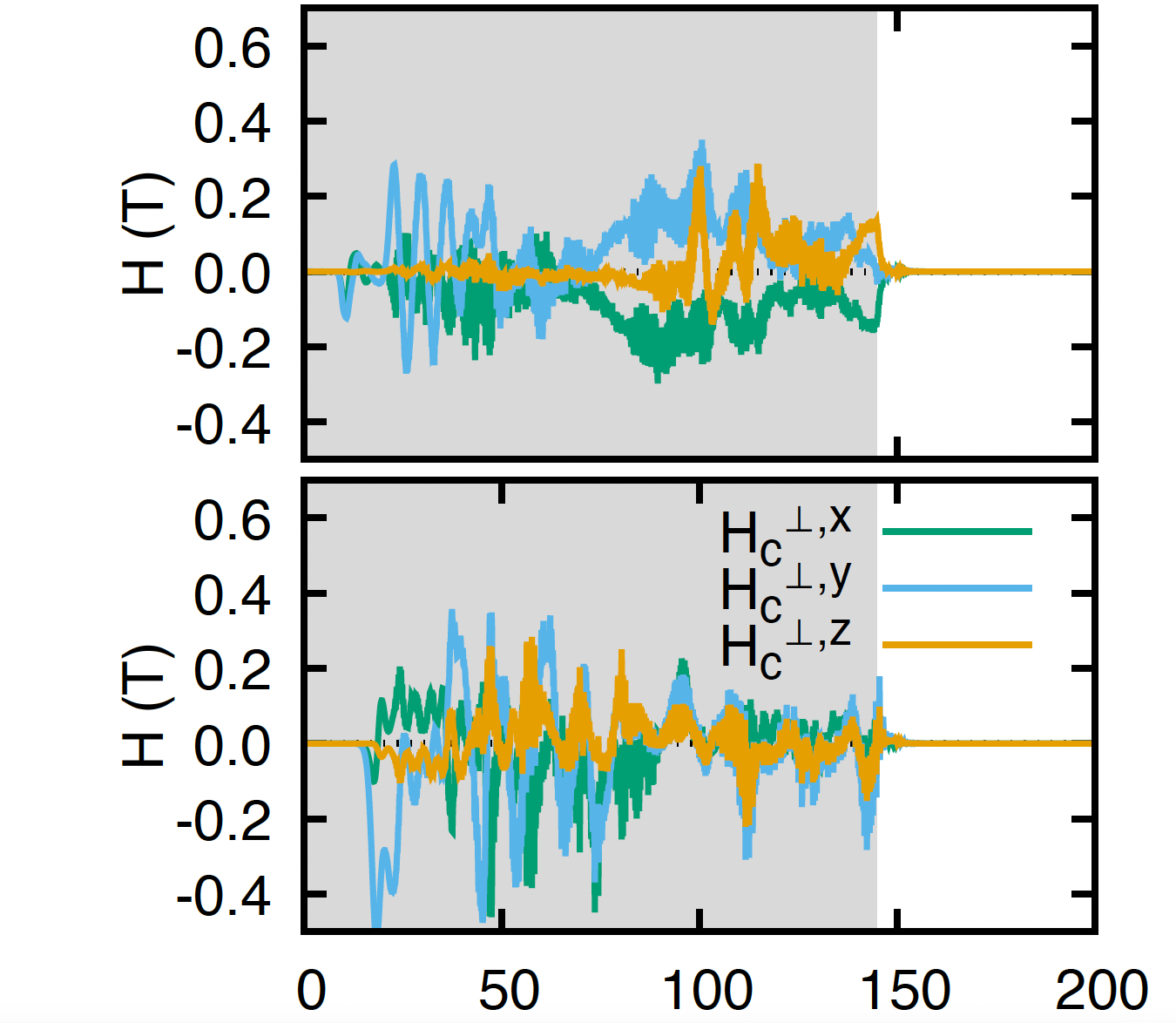}
  \caption{Evolution of the coupling field perpendicular to the direction of magnetisation (xyz components) for the Morse Potential results presented in panel c, d in Fig. \ref{fig::displacements_and_fft}. The frequency corresponding to the THz excitation are 8.6THz (panel a) and 8.4 THz (panel b). \label{fig::perp_comp_field}
}
\end{figure}

\begin{figure}[h]\centering
  \includegraphics[trim={0cm, 4cm, 0cm, 0cm},width=0.85\linewidth]{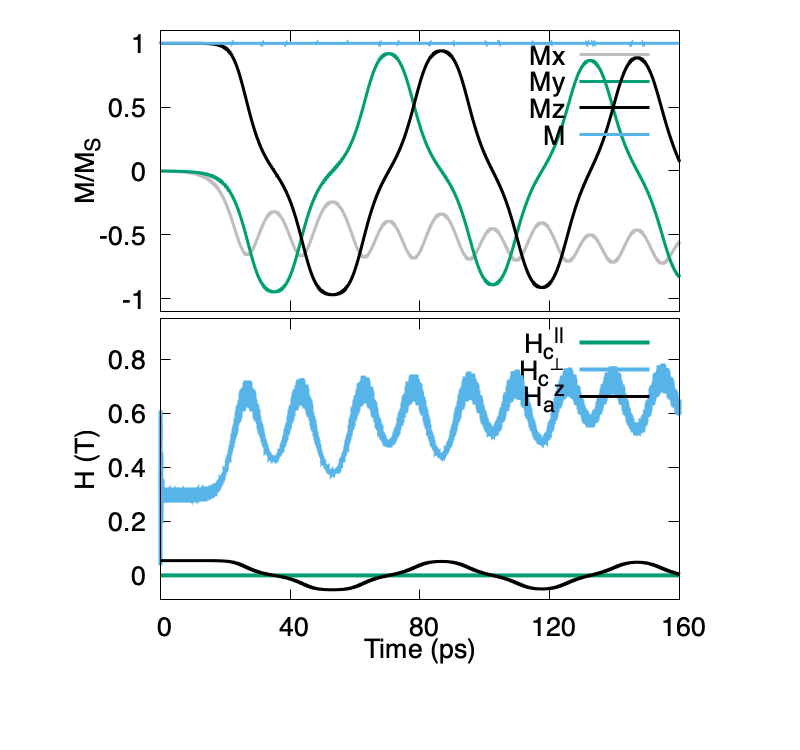}
  \caption{Evolution of magnetisation ($M/M_S$), coupling field parallel and perpendicular to the direction of magnetisation (  $H_c^{\parallel}$ and  $H_c^{\bot}$), anisotropy field ($H_a^z$) for an imposed harmonic variation of the x displacement of the atoms with a frequency of $8.4THz$. \label{fig::res_simple_model}
}
\end{figure}
To show that the developing of the coupling field is solely responsible for the precessional switching, we have imposed on our system a variation of the x displacement with a frequency of $8.4$THz and phase and amplitude as in the numerical simulation. This simple harmonic motion on the x direction leads to precessional switching, as observed in Fig.\ref{fig::res_simple_model}, which is triggered by the magneto-elastic field.
\clearpage

\end{document}